\documentclass[12pt,preprint]{aastex}

\usepackage{emulateapj5}

\newcommand{\tx}[1]{\textrm{#1}}

\newcommand{\kms}{km\,s$^{-1}$}

\newcommand{\be}{\begin{equation}}
\newcommand{\ee}{\end{equation}}

\newenvironment{inlinefigure}{
\def\@captype{figure}
\noindent\begin{minipage}{0.999\linewidth}\begin{center}}
{\end{center}\end{minipage}\smallskip}

\slugcomment{ApJ, 640, 662}

\shorttitle{SLACS II}
\shortauthors{Treu et al.\ }

\begin{document}

\title{The Sloan-Lens ACS Survey II: stellar populations and internal
structure of early-type lens galaxies\altaffilmark{1}}

\author{Tommaso Treu\altaffilmark{2,}\altaffilmark{3}, 
L\'{e}on V. E. Koopmans\altaffilmark{4},
Adam S. Bolton\altaffilmark{5,6},
Scott Burles\altaffilmark{5}
and Leonidas A. Moustakas\altaffilmark{7}}

\altaffiltext{1}{ Based on observations made with the NASA/ESA Hubble
Space Telescope, obtained at the Space Telescope Science Institute,
which is operated by the Association of Universities for Research in
Astronomy, Inc., under NASA contract NAS 5-26555.  These observations
are associated with program \#10174.  Support for program \#10174 was
provided by NASA through a grant from the Space Telescope Science
Institute, which is operated by the Association of Universities for
Research in Astronomy, Inc., under NASA contract NAS 5-26555.}
\altaffiltext{2}{Department of Physics, University of California,
Santa Barbara, CA 93106, USA ({\tt tt@physics.ucsb.edu})}
\altaffiltext{3}{Department of Physics and Astronomy, UCLA, Box
951547, Knudsen Hall, Los Angeles, CA 90095, USA}
\altaffiltext{4}{Kapteyn Institute, P.O. Box 800, 9700AV Groningen,
The Netherlands ({\tt koopmans@astro.rug.nl})}
\altaffiltext{5}{Department of Physics and Kavli Institute for
Astrophysics and Space Research, Massachusetts Institute of
Technology, 77 Massachusetts Ave., Cambridge, MA 02139, USA ({\tt
bolton@alum.mit.edu, burles@mit.edu})} 
\altaffiltext{6}{Harvard-Smithsonian Center for Astrophysics, 60
Garden St., Cambridge, MA 02138, USA ({\tt abolton@cfa.harvard.edu})}
\altaffiltext{7}{Jet Propulsion Laboratory, Caltech, MS169-327,
Pasadena CA 91109 ({\tt leonidas@jpl.nasa.gov})}

\begin{abstract}
We use images taken with the Advanced Camera for Surveys (ACS) on
board the Hubble Space Telescope (HST) to derive effective radii and
effective surface brightnesses of 15 early-type lens galaxies
identified by the Sloan Lens ACS (SLACS) Survey as described in paper
I. The structural parameters are used in combination with stellar
velocity dispersions from the Sloan Digital Sky Survey (SDSS) database
to construct the Fundamental Plane (FP) of lens galaxies. The size of
the sample and the relatively narrow redshift range ($z=0.06-0.33$)
allows us to investigate for the first time the distribution of lens
galaxies in the FP space. After correcting the effective surface
brightnesses for evolution, we find that lens galaxies occupy a subset
of the local FP. The edge-on projection (approximately effective mass
$M$ vs effective mass-to-light ratio $M/L$) is indistinguishable from
that of normal early-type galaxies.  However -- within the fundamental
plane -- the lens galaxies appear to concentrate at the edge of the
region populated by normal early-type galaxies. We show that this is a
result of our selection procedure, which gives higher priority to the
highest velocity dispersions ($\sigma\gtrsim$240\kms). Accounting for
selection effects, the distribution of our galaxies inside the FP is
indistinguishable from that of the parent sample of SDSS galaxies. We
conclude that SLACS lenses are a fair sample of massive (high velocity
dispersion) early-type galaxies. By comparing the central stellar
velocity dispersion ($\sigma$) with the velocity dispersion that best
fits the lensing models ($\sigma_{\rm SIE}$; from paper III) we find
$\langle f_{\rm SIE} \rangle \equiv \langle \sigma/\sigma_{\rm SIE}
\rangle =1.01\pm0.02$ with 0.065 rms scatter. We conclude that within
the Einstein radii (typically $R_e/2$ or $\sim 4$ kpc) the SLACS
lenses are very well approximated by isothermal ellipsoids, requiring
a fine tuning of the stellar and dark matter distribution (bulge-halo
``conspiracy'') in the transition regions of early-type
galaxies. Interpreting the offset from the local FP in terms of
evolution of the stellar mass-to-light ratio, we find for the SLACS
lenses $d \log (M/L_{\rm B}) /dz=-0.69\pm0.08$ (rms 0.11) consistent
with the rate found for field early-type galaxies and with a scenario
where most of the stars were formed at high redshift ($>2$) with
secondary episodes of star formation providing less than $\sim 10$ \%
of the stellar mass below $z=1$.  We discuss star formation history
and structural homogeneity in the context of formation mechanisms such
as collisionless (``dry'') mergers.
\end{abstract}


\section{Introduction}

The properties of early-type galaxies in the local universe obey
several empirical scaling laws such as the correlation between host
velocity dispersion and mass of the central supermassive black hole
(Ferrarese \& Merritt 2000; Gebhardt et al.\ 2000), correlations
between velocity dispersion and stellar ages and chemical composition
(Bender, Burstein \& Faber 1992, 1993), and the correlation between
velocity dispersion, effective radius, and effective surface
brightness known as the Fundamental Plane (FP; Djorgovski \& Davis
1987; Dressler et al.\ 1987). Understanding the origin of these
scaling laws is a challenge for galaxy formation models, because they
imply a degree of homogeneity difficult to explain without invoking a
substantial amount of fine tuning or feedback, still unexplained in
the hierarchical merging scenario.

In particular, the Fundamental Plane can be seen as a scaling relation
between a galaxy's effective (dynamical) mass and effective
(dynamical) mass to light ratio (Faber et al.\ 1987; Bender, Burstein
\& Faber 1992; van Albada, Bertin \& Stiavelli 1995), in the sense
that mass-to-light ratio increases with effective mass (the ``tilt''
of the FP). This tilt could be due (Ciotti, Lanzoni \& Renzini 1996;
Pahre 1998; Trujillo, Burkert \& Bell 2004; Lanzoni et al. 2004) to
trends in stellar populations (more massive galaxies are older and
more metal rich), in the distribution of dark matter (more massive
galaxies contain more dark matter), or in structural properties (the
distribution function depends on mass, e.g. Bertin, Ciotti \& del
Principe 2002). The distribution of galaxies along the FP is also
non-trivial: galaxies do not occupy the whole plane but live in well
defined zones, avoiding a well defined region of the plane (``zone of
avoidance''). From a theoretical point of view it is difficult to
reproduce both the tilt and the distribution of galaxies inside the FP
(e.g. Nipoti, Londrillo \& Ciotti 2003).

In recent years, several groups have measured the FP of early-type
galaxies at cosmological redshift (van Dokkum \& Franx 1996; Kelson et
al. 1997; Pahre 1998; Bender et al.\ 1998; van Dokkum et al.\ 1998;
Treu et al.\ 1999; Kelson et al. 2000; van Dokkum et al. 2001; Treu et
al.\ 2001a,b, 2002; van Dokkum \& Ellis 2003; Gebhardt et al.\ 2003;
van Dokkum \& Stanford 2003; van der Wel et al. 2004; Wujits et al.\
2004; Fritz et al. 2005; Holden et al. 2005; Treu et al. 2005a,b; van
der Wel. et al. 2005; di Serego Alighieri et al.\ 2005; Moran et al.\
2005) to measure the evolution of their mass to light ratio and hence
constrain their star formation history. The results of these studies
can be summarized as follows: massive early-type galaxies (above
10$^{11.5}$ M$_{\odot}$) appear to be evolving slowly below z$\sim$1
consistent with passive evolution of an old stellar population formed
at $z>2$. At smaller masses, signs of recent star formation start to
appear first in the field (at $z\sim0.5$; Treu et al.\ 2002; 2005b,
van der Wel et al.\ 2005) and at higher redshifts possibly even in
clusters (Holden et al.\ 2005). Mass seems to be the dominant
parameter determining the star formation history, while environment
affects $M/L$ only to a lesser degree (Moran et al.\ 2005; Yee et al.\
2005). This trend is consistent with the {\it downsizing} (i.e. star
formation activity moves to lower masses from high to low redshift;
Cowie et al. 1996) scenario seen in a number of studies (e.g. McIntosh
et al.\ 2005; Treu et al.\ 2005a; Juneau et al.\ 2005; see de Lucia et
al.\ 2006 for a theoretical point of view; see Gavazzi 1993 for a
discussion of downsizing {\it ante-litteram} for disk galaxies, based
on fossil evidence). In the case of early-type galaxies, the concept
of downsizing refers to the fact that stars are oldest in the most
massive systems.

Little is known about scaling relations of early-type lens
galaxies. Do they follow the same scaling laws of normal early-type
galaxies? In principle, anomalous structural properties -- such as an
unusually high concentration/mass density -- or mass from a large
scale structure (e.g. a group or a filament) projected along the line
of sight could boost lensing efficiency and therefore a
lensing-selected sample could be biased. Based on the limited amount
of information available so far, no significant difference has been
found between the structural properties (e.g. Treu \& Koopmans 2004)
of lens and non-lens galaxies, supporting the hypothesis that E/S0
lenses are representative of the whole E/S0 population.

In this paper, we exploit the large and homogeneously selected sample
of lenses identified in a relatively narrow redshift range by the
Sloan Lenses ACS Survey (SLACS; Bolton et al.\ 2005; Bolton et
al. 2006, hereafter paper I; www.slacs.org), to study in detail the
Fundamental Plane of lens galaxies, both in terms of tilt and
distribution of galaxies along the plane. We quantify the degree of
homogeneity of the early-type galaxies by measuring the ratio between
stellar velocity dispersion and velocity dispersion of the singular
isothermal ellipsoid (SIE) mass model that best fits the geometry of
the multiple images, using the results of the lens models derived by
Koopmans et al.\ (2006; hereafter paper III).

In addition, we use the sample to study the evolution of the FP of
lens galaxies with redshift as a diagnostic of their stellar
populations. Previous studies on this topic have reached discordant
conclusions. In a pioneering work, Kochanek et al.\ (2000; see also
Rusin et al.\ 2003) assumed isothermal mass density profiles (see
Rusin \& Kochanek 2005 for a more general approach) to convert image
separations into velocity dispersion and construct the FP of
early-type lens galaxies without stellar velocity dispersion
information all the way out to $z\sim1$, finding relatively slow
evolution of their mass-to-light ratio ($d \log (M/L_{\rm B})/dz=
-0.56\pm0.04$) at variance with the faster evolution measured by
direct determinations of the stellar velocity dispersions
(e.g. $-0.72^{+0.07}_{-0.05}\pm0.04$ Treu et al.\ 2005b;
$-0.76\pm0.07$ van der Wel et al. 2005).

Other than a genuine intrinsic difference in the stellar populations,
the origin of this discrepancy could be attributed to departures from
isothermality, to environmental effects, to differences in the
analysis and fitting techniques (e.g. van de Ven, van Dokkum \& Franx
2003 reanalyzed the Rusin et al.\ sample, obtaining faster evolution
$d \log (M/L_B)/dz=-0.62\pm0.13$). Or else, the difference could be
understood in terms of downsizing. In fact -- as a result of the shape
of the luminosity function of early-type galaxies and the scaling of
lensing cross section with mass -- lens samples tend include mostly
massive early-type galaxies ($\sigma\sim250$ \kms) as opposed to
luminosity selected samples which tend to be dominated by galaxies
close to the limiting magnitude.

Early results on the distribution of mass and light in E/S0 lenses
were obtained by the LSD survey (Koopmans \& Treu 2002,2003; Treu \&
Koopmans 2002,2003,2004; hereafter collectively KT), which published
stellar velocity dispersions for 5 lenses in the redshift range
$z\approx0.5-1$. KT found that the image separation under isothermal
assumptions provided a good approximation for stellar velocity
dispersion and that the evolution of stellar mass to light ratio was
in good agreement with that measured for non-lens samples, albeit with
large error bars because of the small sample size. The SLACS sample is
the largest sample of lenses to date with measured stellar velocity
dispersions. It is therefore ideally suited to be combined with the
LSD sample to investigate with higher precision any difference in the
stellar populations of lens and non-lens early-type galaxies, bearing
in mind the SLACS selection process (luminous red galaxies and/or
quiescent spectra, H$\alpha<1.5$\AA; see paper I and Section 2.).

A plan of the paper follows.  Section~2 presents the sample and the
data. Section~3 investigates the internal structure and homogeneity of
SLACS lenses using the FP as a diagnostic tool. Section~4 describes
the redshift evolution of the FP of the combined SLACS+LSD sample and
compares it to that of non-lens early-type galaxies. Section~5
summarizes our results. As in the rest of this series, we adopt Vega
magnitudes and a cosmological model with $\Omega_{\rm m} = 0.3$,
$\Omega_\Lambda = 0.7$, and H$_0 =
70\,h_{70}$\,km\,s$^{-1}$\,Mpc$^{-1}$ (with $h_{70} = 1$ when needed).

\section{Sample Selection and Data Analysis}

\label{sec:data}

The sample analyzed in this paper is composed of the E/S0 lens
galaxies without bright nearby companions identified by the Sloan
Lenses ACS (SLACS) Survey as of March 31 2005 -- the cutoff line for
this first series of papers. A full description of the SLACS Survey
and the selection process -- together with images and spectra of all
the lenses -- is given in paper I of this series (Bolton et al. 2006;
see also Bolton et al. 2004, 2005; see also the SLACS website at
www.slacs.org). For easy reference, we give here a brief
summary. First, lens candidates are found in the SDSS database by
identifying composite spectra made of a quiescent stellar population
and multiple emission lines at a higher redshift (see Bolton et al.\
2004). The spectra are taken from the Luminous Red Galaxies sample
(Eisenstein et al.\ 2001) and the MAIN galaxy sample (Strauss et al.\
2002). Quiescent spectra are selected from the MAIN sample by imposing
a limit on the equivalent width of H$\alpha$ $<1.5$\AA.

Second, the probability of the candidate being a lens as opposed to a
chance overlap within the fiber is computed based on the SDSS stellar
velocity dispersion, the lens and source redshifts, and an isothermal
mass model.  In this model, the probability of being a lens is a
monotonically increasing function of Einstein radius and therefore,
for any given source and lens redshift, of velocity dispersion.

Third, the most promising candidates are followed-up with ACS on HST
in snapshot mode (420s exposures through filters F435W and F814W) to
confirm the lens hypothesis. The initial candidate list for this
proposal consisted of 117 galaxies. The 49 candidates with the largest
Einstein Radius ended up in the SNAPSHOT target list.

In order to facilitate comparison with non-lens sample it is useful to
express our selection criterion -- at least approximately -- in terms
of velocity dispersion instead of Einstein Radius. In general, the
Einstein Radius depends also on the angular distance between the lens
and the source (D$_{ls}$) and between us and the source (D$_{s}$)
since R$_{\rm Einst}\propto \frac{{\rm D}_{ls}}{{\rm D}_{s}}\sigma^2
$.  However, the range in source and lens redshifts is relatively
small for our sample, and therefore the angular distance ratio is
almost constant $\langle \frac{{\rm D}_{ls}}{{\rm D}_{s}}
\rangle$=0.58 with an rms scatter of 0.15. For this reason and for the
dependency of the Einstein Radius on $\sigma^2$, velocity dispersion
is by far the dominant parameter for selection, with lens and source
redshift being a minor second order effect. This is illustrated in
Figure~\ref{fig:histosigma} where we show the distribution of velocity
dispersion for the full list of 117 candidates and the subsample of 49
in the HST target list. A vertical dashed line indicates the threshold
that would be expected for a purely velocity dispersion selected
sample (i.e. the top 42\%=49/117 velocity dispersions). Although the
angular distances ratio softens the cutoff, to first approximation we
can say our sample should be representative of high velocity
dispersion early-type galaxies ($\sigma \gtrsim$ 240 \kms).

As part of our ongoing effort, we successfully proposed a second more
extensive SNAPSHOT program in Cycle 14 (118 targets; GO-10587; PI
Bolton) . The target list for the Cycle-14 program is designed to
provide a uniform distribution of lens velocity dispersions, and
therefore the number of targets increases with decreasing velocity
dispersion to compensate the declining lensing probability. When
completed, the Cycle-14 Survey will allow us to extend our
investigation to early-type galaxies with a larger range of velocity
dispersions.

\begin{inlinefigure}
\begin{center}
\resizebox{\textwidth}{!}{\includegraphics{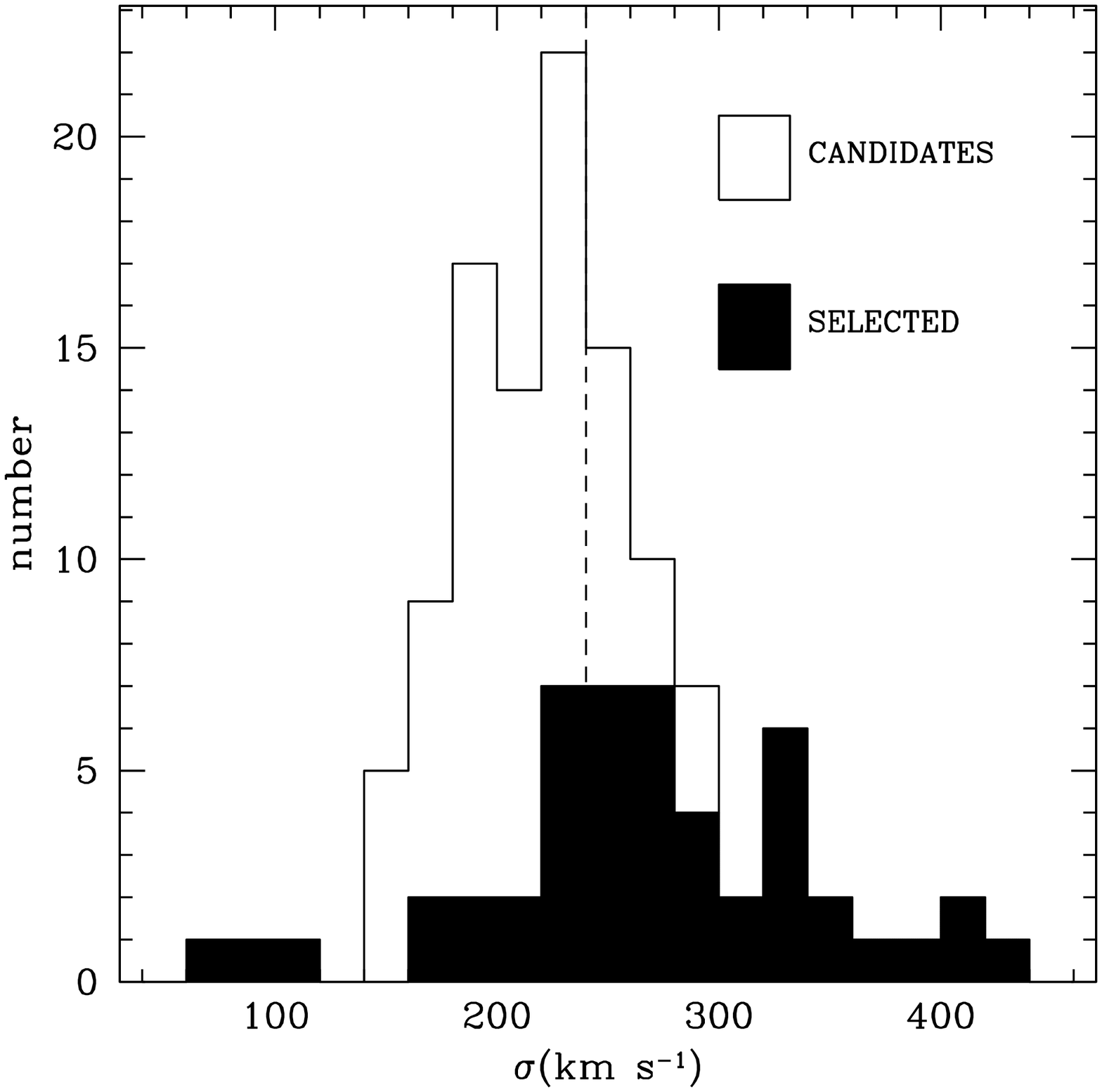}}
\end{center}
\figcaption{Distribution of velocity dispersion for the parent
candidate list and for the 49 targets of the HST snapshot program. The
selection in Einstein Radius picks out the galaxies with largest
velocity dispersion. The vertical dashed line represents the expected
threshold for a velocity dispersion limited sample of 49 objects.
\label{fig:histosigma}}
\end{inlinefigure}

In our treatment of selection effects in \S 3.1, we construct control
samples with exactly the same distribution of $\sigma$. However, we
also model our selection process primarily as a fixed threshold in
$\sigma$, which has the double benefit of being simple and redshift
independent, and of being useful for comparisons with samples of
non-lenses. We show that this approximate selection procedure applied
to the control samples, also reproduces the observed properties of the
SLACS lenses.

Of the 19 lenses confirmed so far, 4 are not considered in this paper:
one lens is rejected because is a spiral galaxy, and the other three
are rejected because they are double lenses or have bright companions
within the SDSS fiber.

Surface photometry was obtained for all the 15 single lens galaxies
characterized by early-type morphology (i.e. E or S0, listed in
Table~\ref{tab:obs}) by fitting de Vaucouleurs profiles to the ACS
images using galfit (Peng et al.\ 2002). Lensed features as well as
neighboring galaxies were carefully masked out during the fit to
minimize contamination.

A Tiny Tim (Krist 2005) generated point spread function (PSF) was used
to convolve the models. The resulting structural parameters --
effective radius ($r_{\rm e}$ and $R_{\rm e}$ respectively in
arcseconds and kpc) and effective surface brightness (SB$_{\rm e}$) --
are fairly insensitive to the choice of PSF as expected because the
effective radii are typically much larger (2-3$''$) than the half
width half maximum of the PSF ($\sim 0\farcs05$). The uncertainty on
the structural parameters was estimated by repeating the 2D fits after
varying the sky levels by twice its standard deviation. As commonly
known (e.g., Hamabe \& Kormendy 1987; J{\o}rgensen, Franx \&
Kj{\ae}rgaard 1993; Treu et al. 2001a), uncertainties on the effective
radius and effective surface brightness are correlated, so that the
combination of parameters that enters the Fundamental Plane (FP$_{\rm
ph}$=$\log R_{\rm e} - 0.32 {\rm SB}_{\rm e}$) is very robustly
determined. Typical errors on FP$_{\rm ph}$ are $\sim$0.02-0.03.

Zeropoints from the latest spectrophotometric calibration of Vega were
adopted (Bohlin \& Gilliland 2004). Magnitudes were corrected for
Galactic Extinction using the SDSS E(B-V) values and the coefficients
A$_{\rm F435W}$=4.325 E(B-V) and A$_{\rm F814W}$=1.984 E(B-V).
Observed radii and magnitudes were transformed into rest frame radii
and surface brightness as described in Treu et al.\ (2001b). As an
independent check of our surface photometry and color transformations,
we used publicly available code (Blanton et al.\ 2003) to compute
absolute magnitudes from the SDSS model photometry finding excellent
agreement (e.g. the difference in absolute B magnitude is less than
0.05 with an rms scatter of 0.15). considering the differences in
spatial resolution and filter system between HST and SDSS.

Stellar velocity dispersions were obtained from the SDSS database. The
measured velocity dispersion within the 3-arcsec diameter SDSS fiber
{$\sigma_{ \rm ap}$} was corrected to a standard central velocity
dispersion (i.e. measured within $r_{\rm e}$/8) using the method
described by J{\o}rgensen, Franx \& Kjaergaard (1995),
i.e. $\sigma$=$\sigma_{\rm ap} \left(\frac{r_{\rm
e}}{8\times1\farcs5}\right)^{-0.04}$. Relevant spectrophotometric
parameters are listed in Tables~1 and~2.

\section{The internal structure of lens galaxies}

In this section, we investigate the internal structure of lens
galaxies by studying their location in the Fundamental Plane space
(\S~3.1), and the ratio between stellar velocity dispersion and
velocity dispersion of the best fitting singular isothermal ellipsoid
as a diagnostic of the mass profile (\S~3.2).

\label{sec:intstuc}

\subsection{The Fundamental Plane of lens galaxies}
\label{ssec:FP}

\noindent
The Fundamental Plane is defined as
\begin{equation}
\label{eq:FP} \log {\rm R}_{\tx{e}} = \alpha \log~\sigma + \beta~\tx{SB}_{\tx{e}} +
\gamma_{\rm FP},
\end{equation} 
($\sigma$ in \kms, R$_{\rm e}$ in kpc, SB$_{\rm e}$ in mag
arcsec$^{-2}$). For this study we adopt as the local relationship the
FP of the Coma cluster in the B band ($\alpha=1.25$, $\beta=0.32$,
$\gamma_{\rm FP}=-9.04$; J{\o}rgensen, Franx \& Kj{\ae}rgaard 1996,
hereafter JFK96, as fitted by Bender et al.\ 1998).  We use Coma as
the local reference for both cluster and field to minimize systematic
uncertainties related to filter transformations, distance
determination, and selection effects; see discussion in Treu et al.\
(2001b), Treu et al.\ (2005b) and van der Wel et al.\ (2005). In the
local universe environmental differences are very small ($<$0.1 mags
arcsec$^{-2}$ at fixed velocity dispersion and effective radius;
Bernardi et al.\ 2003) and negligible for the purpose of this paper.

Figure~\ref{fig:FP} shows the location of the SLACS lens galaxies in
the FP-space, together with the local relationship and Coma galaxies
compiled from J{\o}rgensen, Franx \& Kjaergaard (1992, 1995; hereafter
collectively JFK). The SLACS lens galaxies are distributed parallel to
the local relationship, although on average they are brighter for a
given effective radius and central velocity dispersion, as expected
because of the higher redshift and the consequently younger stellar
populations.  Also, the rms scatter perpendicular to the FP appears to
be larger than that of the Coma sample, because of the relatively
large range in redshifts covered by the SLACS sample. As we will see
in the rest of this section, after correcting for passive evolution,
the scatter is consistent with that observed locally.

\begin{inlinefigure}
\begin{center}
\resizebox{\textwidth}{!}{\includegraphics{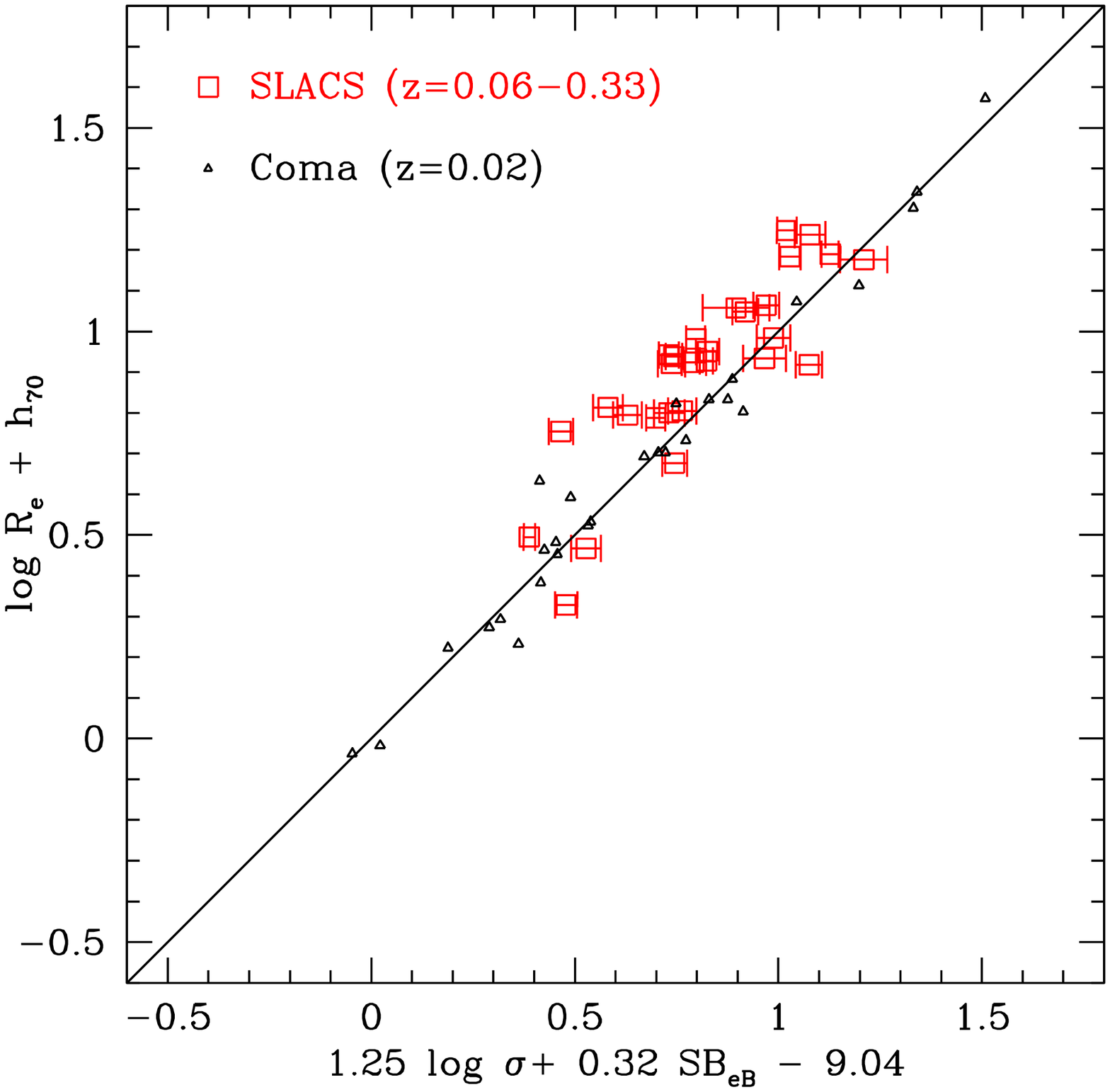}}
\end{center}
\figcaption{Edge-on view of the Fundamental Plane of SLACS E+S0s lens
galaxies. Only error bars due to velocity dispersion are shown since
the error bars due to surface photometry are highly correlated and
move points along the plane. Note that the surface brightness of the
SLACS E+S0s has not been corrected for evolution. The FP of early-type
galaxies in the Coma cluster is shown for comparison (from JFK; black
line and small open triangles).
\label{fig:FP}}
\end{inlinefigure}

To proceed further in our analysis of the FP of lens galaxies, let us
now consider a particularly useful and insightful parametrization of
the FP variables, the so-called $\kappa$-space (Bender, Burstein \&
Faber 1992). The $\kappa$-space is obtained by applying the following
orthogonal\footnote{Considering $2\log \sigma$, $\log I_{\rm e}$ and
$\log R_{\rm e}$ as original coordinates.} transformation to the FP
variables:

\be
\label{eq:kappa1}
\kappa_1 = \frac{2 \log \sigma + \log R_{\rm e}}{\sqrt 2}
\ee

\be
\label{eq:kappa2}
\kappa_2 = \frac{2 \log \sigma + 2\log I_{\rm e} - \log R_{\rm e}}{\sqrt 6}
\ee

\be
\label{eq:kappa3}
\kappa_3 = \frac{2 \log \sigma - \log I_{\rm e} - \log R_{\rm e}}{\sqrt 3} 
\ee

\noindent where I$_{\rm e}$ is the effective surface brightness in
units of L$_{\odot}$ pc$^{-2}$ [e.g. $\log I_{\rm e,B}=-0.4 ({\rm
SB}_{\rm e,B}-27)$ in the B band]. The main advantage of this
transformation is that $\kappa_1$ is proportional to the logarithm of
the effective mass ($M=5\sigma^2 {\rm R}_{\rm e}/G$; Bender, Burstein
\& Faber 1992), while $\kappa_3$ is proportional to the logarithm of
the effective mass to light ratio ($M/L= 5 \sigma^2 {\rm R}_{\rm e}/G
2 \pi I_{\rm e} R_{\rm e}^2$) and therefore the FP is readily
interpreted in terms of physical variables. Furthermore, the plane is
viewed almost perfectly edge-on when projected along $\kappa_3$ and
almost perfectly face-on when projected along $\kappa_2$.

To eliminate evolutionary trends in our study of the distribution of
SLACS lens galaxies in $\kappa$-space, in this section we remove the
average evolution found for field early-type galaxies (Treu et al.\
2005a,b) as $\log I_{\rm e,B,0}=\log I_{\rm e,B} - 0.72 z$ and use
I$_{\rm e,B,0}$ in Equations~\ref{eq:kappa2} and~\ref{eq:kappa3}. Note
that since the redshifts of the SLACS lens galaxies are relatively
small, adopting a different evolutionary rate (e.g. the one from Rusin
et al. 2005) would change the position in $\kappa$-space by a
negligible amount $\sim 0.01-0.02$.

The upper panel of Figure~\ref{fig:kappa} shows the projection of the
SLACS and Coma FP along the $\kappa_2$ axis. It is apparent that the
SLACS lens galaxies follow the same $\kappa_1$-$\kappa_3$ relation of
Coma galaxies, although they occupy preferably the high $\kappa_1$
(i.e. mass) range, as expected since SLACS lens galaxies are selected
to be massive (paper I). A simple least square fit gives
$\kappa_3=(0.21\pm0.02)(\kappa_1-4)+1.03\pm0.01$ for the Coma
galaxies, and $\kappa_3=(0.17\pm0.07)(\kappa_1-4)+1.03\pm0.02$ for the
SLACS galaxies. The two fits are consistent within the uncertainties,
that are larger for the SLACS sample due to its relatively limited
range in $\kappa_1$. The scatter of the SLACS sample is 0.060 in
$\kappa_3$ (0.055 intrinsic after removing in quadrature the average
error on $\kappa_3$), similar to the local intrinsic scatter of 0.05
reported by Bender, Burstein \& Faber (1992).

\begin{inlinefigure}
\begin{center}
\resizebox{\textwidth}{!}{\includegraphics{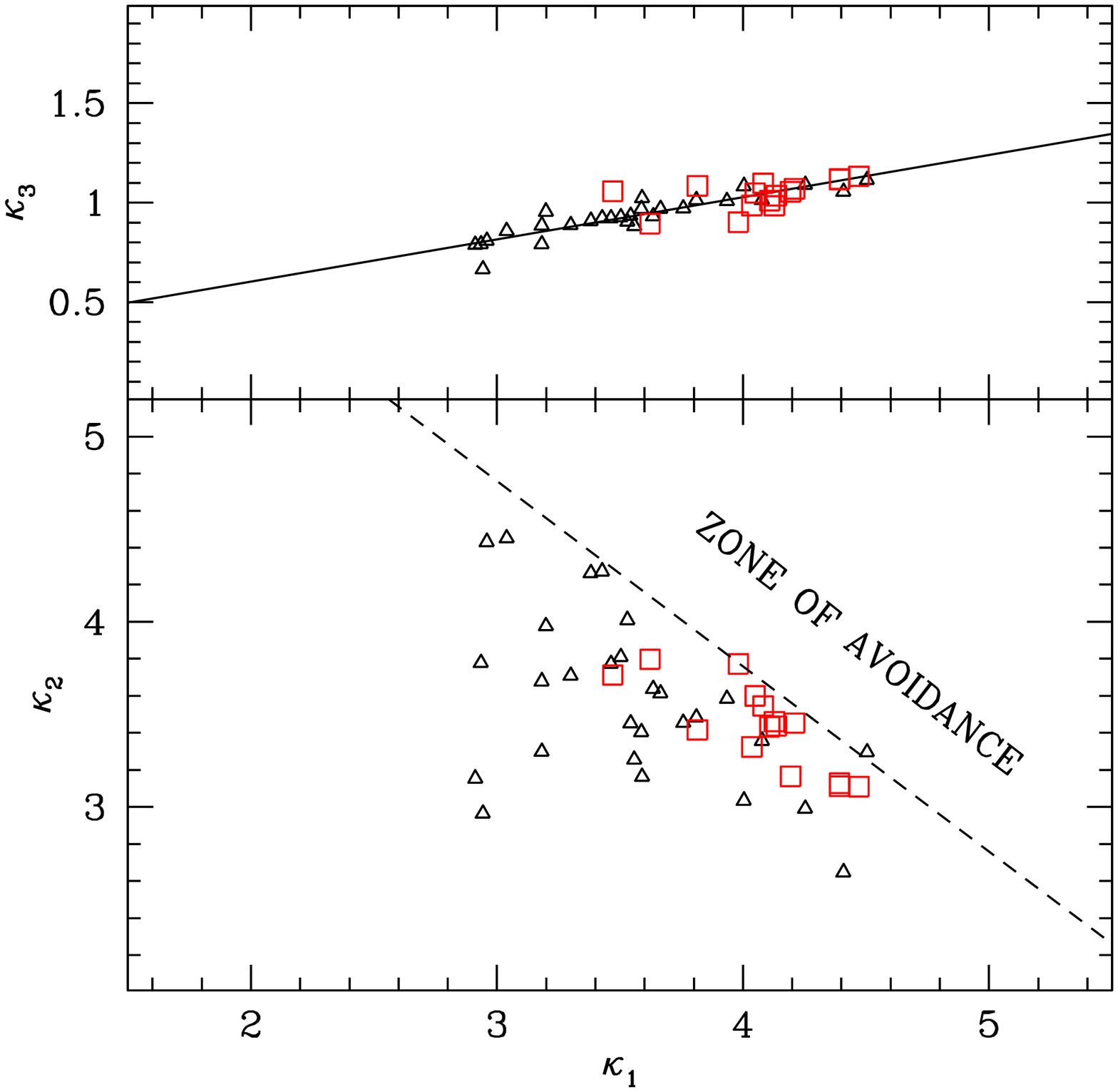}}
\end{center}
\figcaption{Projections of the Fundamental Plane along the so-called
$\kappa$-space variables ($\kappa_1 \propto \log M$, $\kappa_2 \propto
log[(M/L_{\rm B,0}) I_{\rm e,B,0}^3]$ and $\kappa_3 \propto \log
(M/L_{\rm B,0})$). Note that $M/L_{\rm B,0}$ and $I_{\rm e,B,0}$ are
the mass to light ratio and effective surface brightness corrected to
zero redshift according to the evolution measured by Treu et al.\
(2005b). The upper panel shows the projection on the
$\kappa_1-\kappa_3$ plane, corresponding approximately to an edge-on
view of the plane. The bottom panel shows the projection on the
$\kappa_1-\kappa_2$ plane, corresponding approximately to a face-on
view of the plane. SLACS E+S0s lens galaxies are shown as red open
squares. Early-type galaxies in the Coma cluster (from JFK; solid line
and small open triangles) are shown for comparison. The dashed
diagonal line represents the border of the ``zone of avoidance'' not
populated by early-type galaxies (Bender, Burstein \& Faber 1992).
\label{fig:kappa}}
\end{inlinefigure}

Somewhat more surprising is the distribution along the FP, i.e. in the
$\kappa_1$-$\kappa_2$ plane, shown in the bottom panel of
Figure~\ref{fig:kappa}. The SLACS lenses cluster at the edge of the
region normally populated by early-type galaxies, concentrating on the
border of the ``zone of avoidance'' ($\kappa_1+\kappa_2=7.76$; Bender,
Burstein \& Faber 1992). Thus, although lenses have a normal mass
too-light ratio ($\kappa_3$) for a given mass ($\kappa_1$; as shown in
the upper panel of Figure~\ref{fig:kappa}) -- and thus occupy the same
FP as normal galaxies -- they {\it appear} to have higher $\kappa_2$
for a given $\kappa_1$. We show in the rest of this section that this
is the result of our selection procedure. Taking our selection process
into account, lens galaxies are indistinguishable from normal
early-type galaxies.

We can quantify any difference between the SLACS lenses and non-lens
early-type galaxies, by comparing the distribution of distances from
the border of the ``zone of avoidance'', i.e. the distribution of
$\kappa_1 + \kappa_2 - 7.76$, via the Kolmogorov-Smirnov statistic. A
comparison with the JFK Coma sample limited to $\sigma>240$ kms$^{-1}$
(\S 2), suggests that the distributions are indisitinguishable (47\%
probability that they follow the same distribution), once selection
effects are taken into account. However, this comparison is limited to
only 5 objects, and could be affected by differences in the
measurement techniques. Therefore, to draw solid conclusions we have
to consider a larger and more suitable comparison sample.

The best control sample is provided by the parent population of SDSS
selected galaxies. As described in paper I, for each lens, we can
construct a control sample by selecting objects from the same parent
sample (luminous red galaxies and MAIN-sample galaxies from the SDSS),
with the same redshift, and the same limits on spectroscopic
signal-to-noise ratio and strength of H$\alpha$ emission. Since HST
images are not available for this control sample, we use SDSS
photometric and spectroscopic parameters for this comparison, both for
the lenses and the non-lenses. Therefore, we can be sure that there is
no systematic difference in the measured parameters for the lens
sample and the control sample. Also, we can avoid uncertainties
associated with K-corrections by making the comparison in the
observer's frame $\kappa$ space (the o-$\kappa$ space), where
effective radii are expressed in arcseconds and effective surface
brightnesses are not corrected to the rest frame. This is possible
because the control sample for each lens is chosen to lie in a very
thin redshift slice ($\pm$0.005).  For each lens we can then compute
how o-$\kappa_1$+o-$\kappa_2$ ranks within the distribution for the
control objects. If the lenses were a random sample, we would expect
the ranks to be uniformly distributed between 0 and 1. In contrast, we
find that the lenses typically rank amongst the largest values of
o-$\kappa_1$+o-$\kappa_2$, similarly to what is shown in
Figure~\ref{fig:kappa}. Thus -- not accounting for the selection
procedure -- the probability that the lenses are drawn from the
control sample is as low as $7\times10^{-5}$.

Most importantly, however, we have to take into account that the SLACS
lenses are effectively velocity dispersion selected -- and we expect
this selection to skew our lenses toward large values of $\kappa_1$
and $\kappa_2$ (see Equations 2, 3, and 4). To make a proper
comparison, we need to compare the rank of o-$\kappa_1$+o-$\kappa_2$
within the distribution of the parent samples properly selected for
Einstein Radius. Since for each lens Einstein Radius is only a
function of velocity dispersion it is sufficient to cut each parent
sample in velocity dispersion. An important caveat is that the
selection in velocity dispersion cannot be applied simply as as a
symmetric interval around the lens velocity dispersion $\sigma_l\pm N
\delta\sigma_l$ (i.e. N times the error on $\sigma_l$; as was done in
paper I), but requires a more elaborate procedure. This happens
because the lenses are typically on a steeply declining part of the
velocity dispersion function and thus a symmetric interval in $\sigma$
would over-represent galaxies with smaller velocity dispersion. For
this reason, we first select galaxies $\sigma$ in the interval
[$\sigma_l$,$\sigma_l+N\delta\sigma_l$] and then we select the lower
limit so that the total sample contains as many galaxies above and
below $\sigma_l$. The one parameter left $N$, should be chosen to be
the smallest possible that leaves a sizable number of galaxies in each
sample; in the analysis we use N=0.5,1,2 to quantify the effects of
this choice. The effect of this cut in o-$\kappa$ space is illustrated
for a typical system in Figure~\ref{fig:kappasel}. Selecting objects
by velocity dispersion does not significantly alter the edge-on view
of the FP (as expected from Equations 1,2,3,4), but picks out only the
lenses with the highest o-$\kappa_1$+o-$\kappa_2$ (because lenses
typically have the highest velocity dispersion, see \S 2). Including
this effect, the Kolmogorov-Smirnov statistics gives a probability of
12\%,8\%,6\% that that distribution of distances from the zone of
avoidance is the same for the lens and parent samples. In other words,
no significant difference is found once selection effects are taken
into account.

As an additional check, we also consider our approximate selection in
velocity dispersion, by limiting the parent sample of each lens to the
top 42\% velocity dispersions. The effect on each lens is very similar
to the one depicted in Figure~\ref{fig:kappasel}, and the K-S
statistic gives a probability of 27\% that the lens and control sample
are drawn from the same distribution. This confirms that a cut in
velocity dispersion is a useful approximation and thus that our lenses
are representative of early-type galaxies with $\sigma \gtrsim$ 240
\kms.

The analysis of selection effects in velocity dispersion, provides an
interesting clue to the interpretation of the ``zone of
avoidance''. As shown in Figure~4, a slice in $\sigma$ is almost
parallel to the line delimiting the ``zone of avoidance'' (in fact,
combining Equations 1, 2, and 3 yield $\kappa_1$+$\kappa_2$=2.604
$\log \sigma$ + 0.577 $\log I_{\rm e}$-0.119). Therefore, the sharp
cutoff observed in the velocity dispersion function of early-type
galaxies at $\sigma\sim350$ \kms (Sheth et al. 2003) could be one of
the dominant factors to explain the existence of the zone of
avoidance.

For completeness, we investigate possible selection effects related to
the finite aperture of the SDSS fibers. For example, higher surface
brightness galaxies will have a larger fraction of light in the fiber
and thus produce higher signal-to-noise ratio spectra than galaxies
with the same luminosity and larger effective radii. This would make
it easier to measure $\sigma$, and thus finite fiber effects could be
selecting ``compact'' (i.e. small effective radii for a given
luminosity and mass) E/S0, which also occupy the top right portion of
the $\kappa_1$-$\kappa_2$ plane (Equations 3 and 4) . To test this
hypothesis, we compared the distribution of spectroscopic
signal-to-noise ratios for the lenses to that of the control samples,
using the same procedure as above, finding no difference, and thus no
evidence that this is a sizable selection bias. Other mechanisms
related to the finite fiber size -- also weak in magnitude -- are
discussed in paper III.

We conclude that SLACS lenses occupy the same FP as high velocity
dispersion early-type galaxies. It will be interesting to extend this
comparison to lower velocity dispersion galaxies once the Cycle-14
Survey is completed.

\begin{inlinefigure}
\begin{center}
\resizebox{\textwidth}{!}{\includegraphics{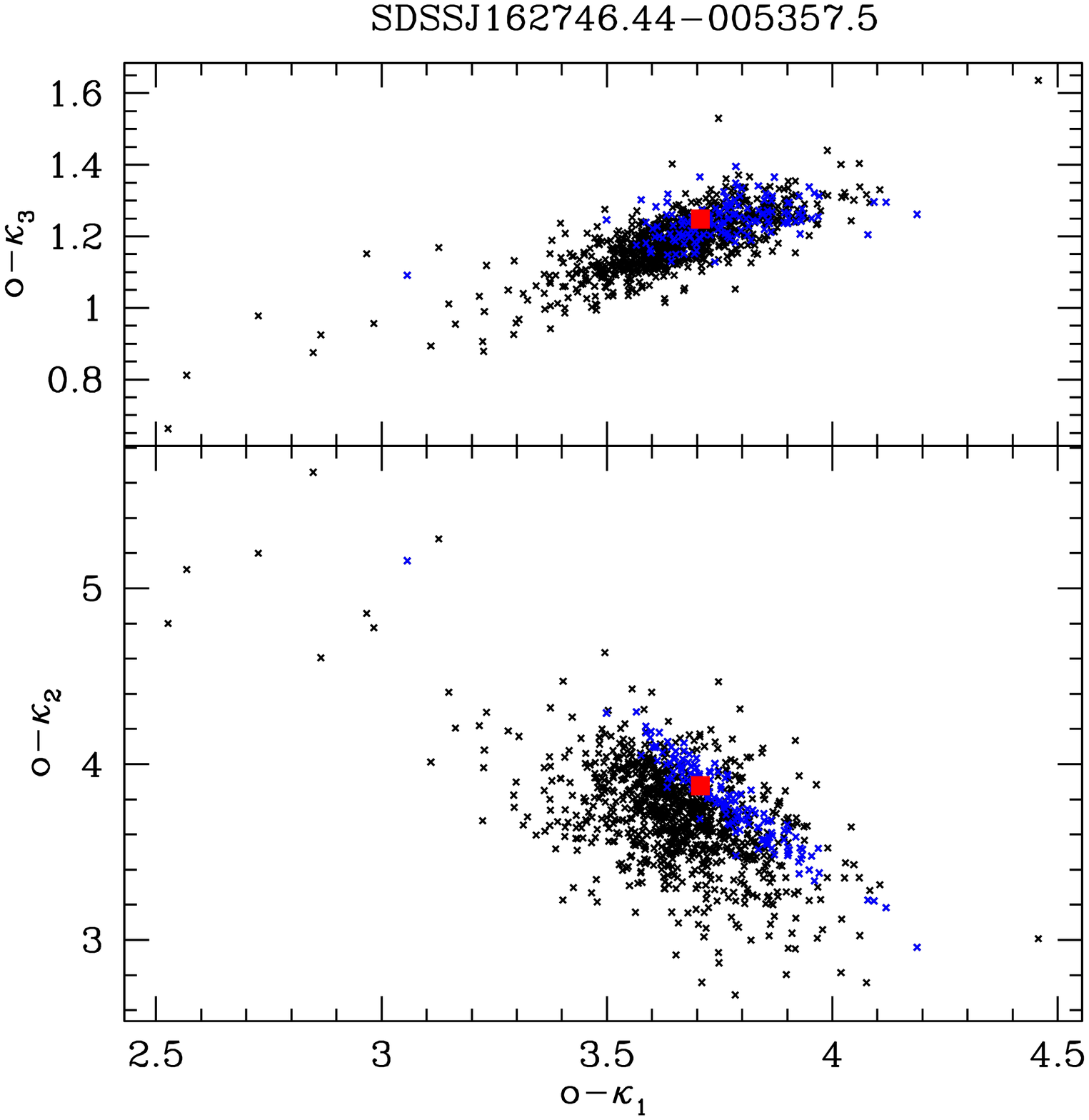}}
\end{center}
\figcaption{Illustration of selection effects in o-$\kappa$ space (the
equivalent of $\kappa$ space in observed quantities, see Section 3.1
for details). For a typical case we show the location of the lens (red
solid square) as well as that of the SDSS control sample (black and
blue crosses). The upper panel shows the projection on the
$\kappa_1-\kappa_3$ plane, corresponding approximately to an edge-on
view of the fundamental plane. The bottom panel shows the projection
on the $\kappa_1-\kappa_2$ plane, corresponding approximately to a
face-on view of the plane. Black crosses represent the full SDSS
parent sample, blue crosses represent the SDSS sub-sample selected as
our own lens sample (i.e. the top 42\% velocity dispersions). Notice
that our selection procedure tends to select objects in the upper
right corner of the o-$\kappa_1$-o-$\kappa_2$ plane because both
variables are monotonically increasing functions of velocity
dispersion $\sigma$.
\label{fig:kappasel}}
\end{inlinefigure}

\subsection{The homogeneity of lens galaxies}

\label{ssec:homo}

A question of great theoretical and practical import is how close
early-type (lens) galaxies are to isothermal density profiles and what
is the intrinsic scatter of their total effective mass density
profiles.  In paper III these questions are addressed via a joint
lensing and dynamical analysis, finding that lenses are extremely
close to isothermal ellipsoids, with effective density profiles well
approximated by power laws $\rho_{\rm tot}\propto r^{-\gamma'}$ with
$\langle \gamma' \rangle =2.01^{+0.02}_{-0.03}\pm0.05$ and an
intrinsic rms scatter of 0.12 in $\gamma'$.

In this paper we take an independent and empirical approach and
quantify the homogeneity of early-type galaxies and departures from
isothermality by studying the ratio between stellar velocity
dispersion and velocity dispersion of the best fitting singular
isothermal ellipsoid $\sigma_{\rm SIE}$ (paper III). The ratio $f_{\rm
SIE}=\sigma/\sigma_{\rm SIE}$ is shown in Figure~\ref{fig:fSIE}. Both
the aperture velocity dispersion and the central velocity dispersion
agree remarkably well with $\sigma_{\rm SIE}$. In both cases the
average ratio is very close to unity ($\langle f_{\rm SIE} \rangle
=1.010\pm0.017$ and $0.950\pm0.015$) with a remarkably small scatter
(0.065 and 0.055 respectively). No significant correlation is found
between $f_{\rm SIE}$ and mass indicators (i.e. $\sigma$ or
$\kappa_1$).

\begin{inlinefigure}
\begin{center}
\resizebox{\textwidth}{!}{\includegraphics{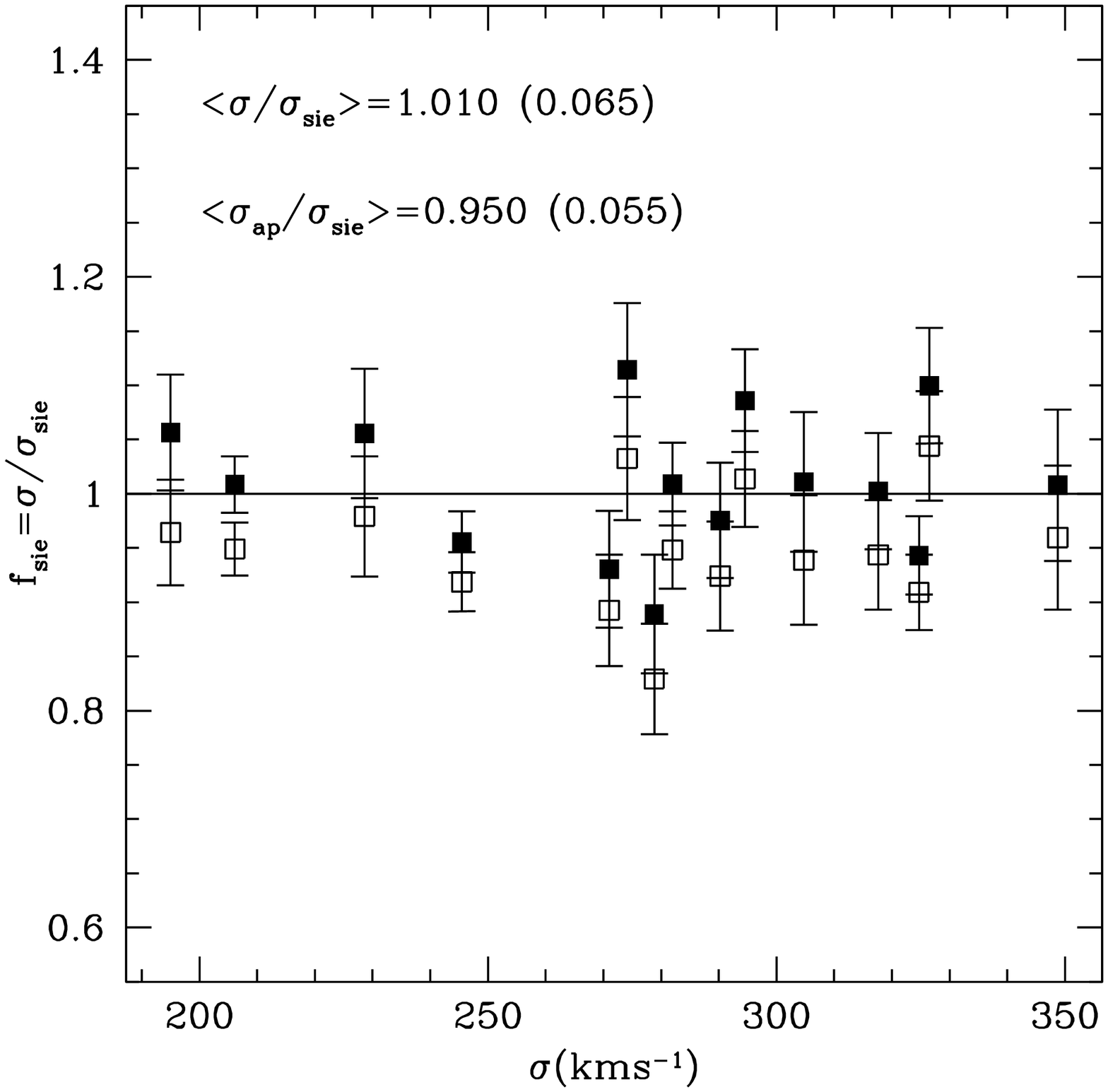}}
\end{center}
\figcaption{Ratio between stellar velocity dispersion and velocity
dispersion of the singular isothermal ellipsoid that fits best the
lensing geometry. Solid points represent $\sigma$ the central velocity
dispersion, i.e. corrected to a circular aperture of radius $r_{\rm
e}/8$. Open points represent $\sigma_{\rm ap}$ the stellar velocity
dispersion as measured within the aperture of the SDSS fibers (3''
diameter).
\label{fig:fSIE}}
\end{inlinefigure}

From the LSD sample of five higher redshift lenses KT found
$\langle$f$_{\rm SIE}\rangle=0.87\pm0.04$ with an rms scatter of 0.08
(including non-LSD lenses PG1115+080 and B1608+656 raises the average
to $0.95\pm0.07$, rms 0.17; Treu \& Koopmans 2002b; Koopmans et al.\
2003). Although the LSD value is also close to unity, the two measured
ratios for LSD and SLACS are only marginally consistent (depending on
whether PG1115+080 and B1608+656 are included or not), perhaps
indicative of significant intrinsic scatter or redshift dependent
average (c.f. paper III).  Similarly, van de Ven et al.\ (2003)
measured the ratio $\sigma / \sigma_c$ for a sample of 7 lenses (4 in
common with TK04, plus two additional E/S0 lenses and the bulge of the
barred spiral Q2237+030; Huchra et al. 1985). Their parameter
$\sigma_c$ is derived from $\sigma_{\rm SIE}$ by solving the spherical
Jeans equation and assuming isothermal total mass distribution and a
Hernquist (1990) luminous mass profile. They found $\sigma /
\sigma_c=1.07$ with an rms scatter of 0.27, corresponding to $f_{\rm
SIE}=1.07 \times g $ with g=0.9-1.01 (i.e $f_{\rm SIE}=0.96-1.08$) for
a range of anisotropy parameters (see van de Ven et al. 2003 for
details).

Redshift dependent effects could be present if contamination from a
group/cluster environment or from large scale structures was an
important source of bias. The importance of this effect has been
estimated by Keeton \& Zabludoff (2004) who analyze mock realizations
of groups like PG1115+080 and find that external convergence leads to
an of overestimate $\sigma_{\rm SIE}$ by $\sim 6$\% with a broad tail
extending up to 10-15\% (for additional discussion of environmental
effects see Holder \& Schechter 2003 and Dalal \& Watson 2004 for a
theoretical point of view, and Fassnacht \& Lubin 2002 and Fassnacht
et al.\ 2005 for an observational point of view). In fact, for a fixed
velocity dispersion (e.g. 250 \kms) the Einstein Radius in arcseconds
is approximately the same ($1\farcs135$) for the typical SLACS
($z_l=0.2$, $z_s=0.6$) and LSD redshifts ($z_l=0.5$, $z_s=2.0$),
resulting in a larger Einstein Radius in kpc for LSD lenses
(approximately double). Thus, for a large scale structure of fixed
projected density, the mass contribution within the Einstein Radius
would be $\sim4$ times larger, while the lens mass within the Einstein
Radius would only grow by a factor of $\sim2$.

Based on the SLACS and LSD samples we conclude that on average the
approximation $\sigma=\sigma_{\rm SIE}$ appears to be working
surprisingly well, due perhaps to some yet unexplained mechanism that
couples stellar and dark mass, a bulge-halo ``conspiracy'' similar to
the disk-halo conspiracy of spiral galaxies. Significant departures
are seen however in individual cases (up to 11\% for SLACS and 30\%
for LSD + PG1115+080 + B1608+656, including observational errors; see
also Kochanek et al.\ 2006), sufficient to introduce sizable
uncertainties in applications that require a precise radial dependency
of the mass model (e.g. determination of the Hubble Constant from
gravitational time delays; note however that time delays are sensitive
to the slope in the region of the images -- Kochanek 2002 --, while
here we are considering the average slope inside the largest of the
effective and Einstein radii) as discussed in paper III.

\section{The evolution of the Fundamental Plane}

\label{sec:FPev}

Under appropriate assumptions (e.g. Treu et al.\ 2001b, 2005b), the
evolution of the Fundamental Plane can be interpreted as general
trends in luminosity evolution of the stellar populations.  If
$\sigma$ and R$_{\rm e}$ do not evolve with redshift, for an
individual galaxy (labeled by the superscript $i$)
\be 
\gamma_{\rm FP}^i\equiv \log R_{\tx{e}}^i - \alpha \log \sigma^i - \beta \tx{SB}_{\tx{e}}^i,
\label{eq:gammai}
\ee 
the offset with respect to the prediction of the FP ($\Delta
\gamma_{\rm FP}^i \equiv \gamma_{\rm FP}^i-\gamma_{\rm FP}$) is
related to the offset of the $M/L$ by
\be
\Delta \log \left( \frac {M}{L} \right)^i = -\frac{\Delta \gamma_{\rm FP}^i}{2.5 \beta},
\label{eq:dgdml}
\ee
which can be used to measure the average evolution and/or scatter of
$M/L$ at given $M$ (see extended discussions in Treu et al.\ 2001b and
Treu et al.\ 2005b).  As usually done in FP studies, the evolution of
the effective M/L can be connected to the star formation history by
assuming that $\Delta \log (M/L) = \Delta \log (M_*/L$) (i.e stellar mass
$M_*\propto M$).

The derived evolution of the effective mass to light ratio of SLACS
lens galaxies is shown in Figure~\ref{fig:FPBlsd} (red solid
squares). Although the redshift range covered by the SLACS sample is
relatively small, there is a clear indication of evolution, as
expected for evolving stellar populations. A simple least square fit
to the evolutionary rate, gives $d\log (M/L_{\rm
B})/dz$=$-0.69\pm0.08$ with an rms scatter of the residuals of 0.11 in
$\Delta \log (M/L_{\rm B})$. Including the higher redshift lens
galaxies from LSD gives $d\log (M/L_{\rm B})/dz=-0.76\pm0.03$, leaving
unchanged the rms scatter.

\begin{inlinefigure}
\begin{center}
\resizebox{\textwidth}{!}{\includegraphics{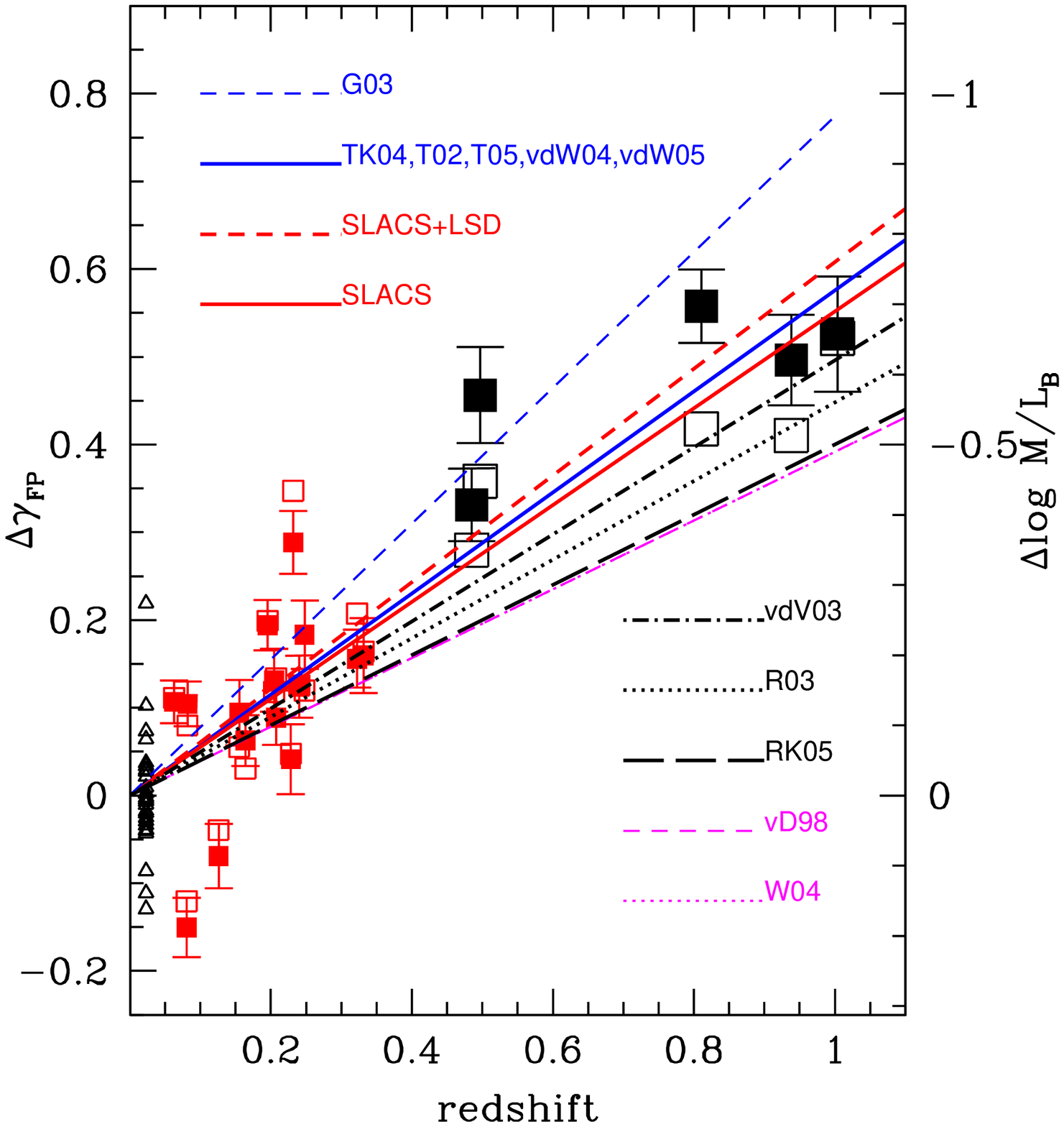}}
\end{center}
\figcaption{Evolution of the effective mass-to light ratio of
early-type galaxies from the evolution of the FP. The small solid
(open) red squares represent the evolution of SLACS lenses computed
using $\sigma$ ($\sigma_{\rm SIE}$) in Equation~\ref{eq:gammai}. The
large solid black squares represent LSD galaxies at $z>0.4$. The small
open triangles at $z=0.023$ represent the sample of Coma E/S0 from
JFK.  Evolutionary trends for a number of recent determinations are
also plotted as solid, dotted and dashed lines. The thick red lines
represent the best linear fit to the SLACS data (solid) and to the
SLACS+LSD data (dashed).  The thick blue line summarizes the very
similar values found by Treu \& Koopmans (2004; TK04), Treu et al.\
(2002, 2005a,b; T02, T05), van der Wel et al.\ (2004, 2005;
vdW04,05). The thick black lines represent measurements for lens
samples by van de Ven (2003 vdV03; dotted-short dashed), Rusin et al.\
(2003, R03; dotted) and Rusin \& Kochanek (2005; RK05; dashed). The
thin blue line represent field measurements from Gebhardt et al.\
(2003; G03; dashed). The thin magenta lines represent cluster
measurements from van Dokkum et al.\ (1998; vD98; dashed) and Wuyts et
al.\ 2004 (W04; dotted). All the lines are based on spectroscopic
stellar velocity dispersions, except for vdV03, R03, RK05.
\label{fig:FPBlsd}}
\end{inlinefigure}

For our sample of lenses we can check the effect of adopting
$\sigma_{\rm SIE}$ instead of $\sigma$ in Equation~\ref{eq:gammai}
(Kochanek et al.\ 2000). The resulting values of $\Delta \log
(M/L_{\rm B})$ are plotted in Figure~\ref{fig:FPBlsd} as open symbols
showing that in general the change is smaller than the error
bars. Adopting $\sigma_{\rm SIE}$ in Equation~\ref{eq:gammai} changes
the linear fit to the evolution of the $M/L_{\rm B}$ of the SLACS
lenses to $d\log (M/L_{\rm B})/dz$=$-0.73\pm0.08$, i.e. less than the
uncertainty. The change is somewhat larger for the LSD lenses, perhaps
due to the larger relative importance of the lens environment or large
scale structure along the line of sight, as discussed in
\S~3.2. Extending the fit based on $\sigma_{\rm SIE}$ to the SLACS+LSD
lenses yields $d\log (M/L_{\rm B})/dz$=$-0.66\pm0.02$.

\subsection{Discussion and comparison with previous work}

\label{ssec:previous}

In Figure~\ref{fig:FPBlsd} the evolution measured for the SLACS and
LSD lenses is compared to the best fit values found by previous
studies (see caption to Figure~\ref{fig:FPBlsd} for key to lines) . We
emphasize that such comparisons should be taken with caution given the
different selection criteria for the various samples and the
importance of accounting for selection effects to interpret the
evolution of the FP (see discussions in Treu et al.\ 2001b, 2002,
2005b).

\begin{inlinefigure}
\begin{center}
\resizebox{\textwidth}{!}{\includegraphics{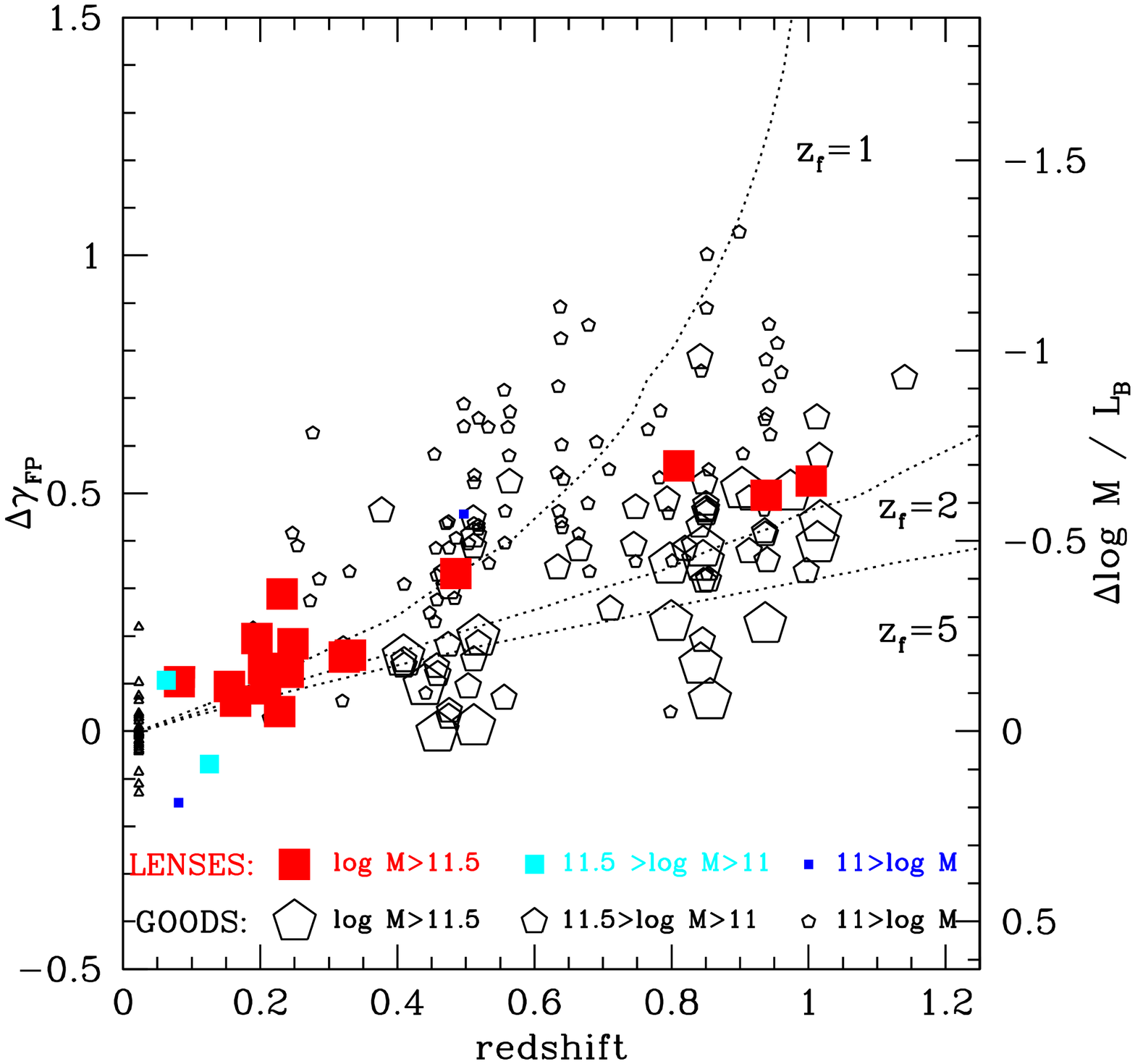}}
\end{center}
\figcaption{Comparison between the evolution of $M/L_{\rm B}$ for
SLACS and LSD lenses (solid colored symbols) and that for field
early-type galaxies in the GOODS-N field (from Treu et al.\ 2005a,b;
open black symbols). The size of the symbols encodes the effective
mass (in solar units). The local reference sample of the Coma Cluster
is also shown as small open symbols at $z=0.023$. Tracks corresponding
to passive evolution of a single burst stellar population formed at
$z=1,2,5$ are overlaid for guidance (see Treu et al.\ 2005b for a
description of the models). Note that most lenses are in the highest
effective mass bin and evolution of the GOODS-N field lenses depends
strongly on the effective mass.
\label{fig:FPGOODS}}
\end{inlinefigure}

Within the error bars, the SLACS+LSD evolution in $M/L$ agrees with
the average evolutionary rate found for non-lens field early-type
galaxies by the most recent and comprehensive studies.  In comparison
with studies of other samples of lens galaxies, however, the SLACS+LSD
sample yields a faster evolutionary rate than the Rusin et al.\
(2003;$-0.56\pm0.04$) and Rusin \& Kochanek (2005; $-0.50\pm0.19$)
studies\footnote{A measure of the systematic uncertainties related to
the methodology is given by the van de Ven et al.\ (2003) re-analysis
of the same sample of lenses, which gives $-0.62\pm0.13$.}. The
discrepancy is reduced if only SLACS lenses are considered, although
it must be kept in mind that SLACS lenses are selected to be red and
quiescent and therefore biased toward slower evolution. Alternatively,
the discrepancy is reduced if $\sigma_{\rm SIE}$ is adopted for the
SLACS+LSD lenses, perhaps indicating that part of the reason for this
mild discrepancy lies in the isothermal approximation for the highest
redshift systems.

In conclusion, noting the various selection effects and still
relatively large uncertainties and small numbers at high-z for the
lens samples, no convincing evidence is found that the evolution of
the Fundamental Plane is different for lens and non-lens E/S0
galaxies. Possible differences are best investigated in the context of
the mass dependency ('downsizing') reported for non-lens samples. This
is illustrated in Figure~\ref{fig:FPGOODS}, where we compare the
evolution of the FP for SLACS+LSD lenses to that found by Treu et al.\
(2005a,b) for a sample of 141 field E/S0 galaxies in the GOODS-N
field. The vast majority of the lenses have masses above
$10^{11.5}$M$_{\odot}$ and follow approximately the evolutionary path
observed for field E/S0 in the same mass range, i.e. passive evolution
of an old stellar population formed at $z\sim2$ with limited amounts
of stars formed in secondary bursts at lower redshift. In order to
avoid duplications with previous work we do not discuss this in more
detail or more quantitatively, referring instead to the extended
discussion and analysis in Treu et al.\ (2005b).

A possible hint of difference is found when comparing the $M/L$ of the
three high redshift LSD lenses to that of the high-mass E/S0 in
GOODS. The three lenses are in the lower $M/L$ part of the
distribution. According to a K-S test, the probability that the three
LSD lenses are drawn from the same distribution as that of the massive
E/S0 is 6\%. The lowest $M/L$ of the three lenses is HST1417, which
has been known for quite some time to show fast evolution (Ohyama et
al.\ 2002; Gebhardt et al.\ 2003; Treu \& Koopmans 2004; van der Wel
et al.\ 2004) and happens to be the least massive of the three. This
possible difference must be taken with great care given the extremely
different selection criteria and their importance in interpreting the
FP (e.g. Treu et al. 2005a,b). In particular, the spectroscopic limit
for LSD was not as deep as for the GOODS study, possibly imposing a
more stringent upper limit to the measurable $M/L$. Given the small
number statistics involved, we do not attempt to quantify the
significance of this difference any further, but we await the
collection of a larger sample (suitable targets are rapidly becoming
available, e.g. Cabanac et al. 2005). We conclude by emphasizing that
lenses occupy the same FP as normal E/S0 (\S 3.1) and they have
exactly the same $M/L$ for a given $M$ (top panel of
Figure~3). Therefore, their distribution inside the FP and any
selection in $\sigma$ is irrelevant to the determination of the
evolution of $M/L$.

\section{Summary and conclusions}

\label{sec:sum}

We have performed 2D surface photometry on the HST-ACS images of 15
lens early-type galaxies identified by SLACS. As described in paper I
(Bolton et al.\ 2006) the lenses are selected from the luminous red
galaxy and main galaxy sample of the SDSS database with quiescent
spectra, i.e. equivalent width of H$\alpha<1.5$\AA. In combination
with stellar velocity dispersions from the SDSS database, we have
constructed the Fundamental Plane of lens galaxies and measured its
evolution with redshift. We have compared the measured stellar
velocity dispersion with the velocity dispersion of the singular
isothermal ellipsoid that best fits the lensing geometry (c.f. paper
III) to study the homogeneity of lens galaxies and the accuracy of the
isothermal approximation to measure the evolution of the FP of lens
galaxies. The main results can be summarized as follows:

\begin{enumerate}

\item SLACS lenses define a Fundamental Plane correlation over almost
a decade in effective radii. The lenses are typically brighter than
local early-type galaxies for a given velocity dispersion and
effective radius, consistent with lower mass to light ratios,
i.e. younger stellar populations at $z=0.06-0.33$ than today.

\item After correction for evolution of the stellar populations, the
SLACS lenses fall on the FP of early-type galaxies in the local
Universe.  The edge-on projection of the FP of SLACS lenses is
consistent with that of local galaxies within the errors. In contrast,
SLACS lenses occupy a relatively small portion of the plane,
concentrating along the border of the so-called ``zone of avoidance''
of local early-type galaxies. We show that this is the result of our
selection procedure focused on the systems with the highest velocity
dispersion. Accounting for the selection procedure, the distribution
of distances from the ``zone of avoidance'' is indistinguishable from
that of the SDSS parent samples (MAIN and LRG). We conclude that the
SLACS lenses are a fair sample of high velocity dispersion ($\sigma
\gtrsim 240$ \kms) early-type galaxies.

\item The ratio between the central stellar velocity dispersion
($\sigma$) and velocity dispersion of the singular isothermal
ellipsoid ($\sigma_{\rm SIE}$) that best fits the lensing geometry is
found to be $\langle f_{\rm SIE} \rangle =1.010\pm0.017$ with an rms
scatter of 0.065.  The isothermal approximation for the SLACS lenses
works better than that for the LSD sample of 5 galaxies at higher
redshifts ($\langle f_{\rm SIE} \rangle =0.87$ with rms scatter
0.08). If this redshift dependency is confirmed by larger and
homogeneously selected samples at higher redshift, possible
explanations include an intrinsic change in the properties of
early-type galaxies with cosmic time or simply by an increased
contribution of external convergence, resulting from group or clusters
associated with the lens or large scale structures along the line of
sight.

\item Interpreting the evolution of the FP in terms of evolution of
the stellar populations, the effective mass to light ratio of SLACS
lenses evolves as $d \log (M/L_{\rm B}) / dz = -0.69\pm0.08$ with an
rms scatter of 0.11. Adding the 5 galaxies from the LSD sample, the
best fit evolutionary rate changes to $-0.76\pm0.03$ leaving the
scatter unchanged. The evolutionary rate changes within the error if
$\sigma_{\rm SIE}$ is used to construct the FP instead of $\sigma$.

\end{enumerate}

We now briefly discuss these results in terms of their implications
for our understanding of the formation and evolution of early-type
galaxies\footnote{To avoid duplications within the series, the main
discussion is left for paper III.}.  From the point of view of stellar
populations, we find that the SLACS lenses have mostly old stellar
population, with at most a small ($<10$\%) contribution of stellar
mass accreted at $z<1$. This is in agreement with what is found
(e.g. Treu et al. 2005a,b) for non-lens early-type galaxies of
comparable mass ($\langle\sigma \rangle=279$, r.m.s. scatter 45 km
s$^{-1}$). This conclusion is unlikely to be significantly biased by
our selection criterion against H$\alpha$ emission, since at the
redshifts of the current SLACS sample ($\langle z \rangle $=0.19),
emission lines are not frequent even in morphologically selected
samples (e.g. Treu et al.\ 2002).

Concluding that most stars are old, however, does not answer the
question of how and when the stars and dark matter are assembled. An
often-invoked mechanism for the assembly of massive early-type
galaxies (e.g. Khochfar \& Burkert 2003) are the so-called ``dry''
mergers, i.e. mergers that do not involve a significant amount of cold
gas and hence not associated with star formation (see Bell et al.\
2006 and van Dokkum 2005 for recent results and discussions). Since
the stars in merging galaxies are already old and the dynamical
timescales for mergers are rather short (a few hundred million years),
dry mergers provide an efficient way to assemble large amounts of old
stars with little remnants.

However, the old stellar ages are only one of the observational tests
of the dry mergers hypothesis. The tight scatter of empirical scaling
laws such as the Fundamental Plane and the black-hole mass $\sigma$
relation must also be explained under this scenario. Numerical
simulations show that plausible configurations of major dry mergers
preserve the tightness of the FP, i.e. if two galaxies start on the FP
they also end up on the FP (Nipoti, Londrillo \& Ciotti 2003;
Gonzalez-Garcia \& van Albada 2003; Boylan-Kolchin, Ma \& Quataert
2005), with an edge-on thickness comparable to that observed. In
contrast, other scaling laws such as the Kormendy and Faber-Jackson
relationships and the black-hole mass $\sigma$ relationship, are not
naturally preserved by dry mergers (Nipoti, Londrillo \& Ciotti
2003). A substantial amount of dissipation seems to be necessary to
preserve those (e.g. Kazantzidis et al. 2005).

The observations present here provide a new series of tests for this
hypothesis. First, we measure the distribution of early-type galaxies
within the FP: this also should be reproduced by a successful
formation mechanism. At the moment there is no good explanation for
the ``zone of avoidance'', nor does it appear clear whether dry
mergers alone can populate the right portion of the plane (as
suggested perhaps by the difficulties in reproducing the Kormendy and
Faber-Jackson relation). It is possible that dissipation may end up to
be necessary to move the objects towards higher concentration, against
the boundary of the ``zone of avoidance'' (see Bender, Burstein \&
Faber 1992).

Second, the proposed mechanism must also explain the distribution of
mass {\it inside} each galaxy. Our study shows very clearly that, at
these scales (typically $R_{\rm e}/2$), early-type lens galaxies are
well approximated by singular isothermal ellipsoids (see also KT and
paper III). Not only this mass density profiles differs significantly
from cosmologically motivated ones (e.g. Navarro et al.\ 1997; Moore
et al.\ 1998), but also it requires a significant amount of fine
tuning between the distribution of luminous and dark matter
(bulge-halo ``conspiracy''). It has been suggested that the total mass
density profile can act as a ``dynamical attractor'' for collisionless
particles (i.e. stars and dark matter; Loeb \& Peebles 2003; Gao et
al. 2004) in a similar way to the close-to-isothermal profiles
obtained for (incomplete) violent relaxation scenarios (e.g. van
Albada 1982). Our measurements provide a quantitative test for this
idea. From the point of view of dry mergers, the close-to-isothermal
mass density profile raises two problems. The first is whether such
property is preserved during dry mergers. Assuming that this is the
case, the second problem is how did the progenitors get their initial
mass density profile. Simulations of dry mergers typically start with
input galaxies already ``dense'' and close-to-isothermal, but -- as we
argue in paper III -- this process cannot be traced back {\it ad
infinitum}. A different process -- perhaps gas rich mergers -- --
appear to be needed at some point to create these high concentration
objects. This brings us back to the main unanswered question: what is
the origin of the bulge-halo ``conspiracy''?

\acknowledgments

We thank the referee for a careful and insightful report that
significantly improved the paper. We are grateful to Luca Ciotti,
Barbara Lanzoni, Simon White for useful conversations.  We acknowledge
financial support from HST grants (STScI-AR-09222; STScI-GO-10174). TT
acknowledges support from NASA through Hubble Fellowship grant
HF-01167.1 and UCLA for being such a welcoming and stimulating Hubble
Fellowship host institution during the initial phases of this project.
The work of LAM was carried out at the Jet Propulsion Laboratory,
California Institute of Technology, under a contract with NASA.  This
project would not have been feasible without the extensive and
accurate database provided by the Digital Sloan Sky Survey (SDSS).
Funding for the creation and distribution of the SDSS Archive has been
provided by the Alfred P. Sloan Foundation, the Participating
Institutions, the National Aeronautics and Space Administration, the
National Science Foundation, the U.S. Department of Energy, the
Japanese Monbukagakusho, and the Max Planck Society. The SDSS Web site
is http://www.sdss.org/.  The SDSS is managed by the Astrophysical
Research Consortium (ARC) for the Participating Institutions. The
Participating Institutions are The University of Chicago, Fermilab,
the Institute for Advanced Study, the Japan Participation Group, The
Johns Hopkins University, the Korean Scientist Group, Los Alamos
National Laboratory, the Max-Planck-Institute for Astronomy (MPIA),
the Max-Planck-Institute for Astrophysics (MPA), New Mexico State
University, University of Pittsburgh, University of Portsmouth,
Princeton University, the United States Naval Observatory, and the
University of Washington.

\clearpage

\clearpage
\begin{deluxetable}{lccccccc}
\tablecaption{Summary of relevant measurements \label{tab:obs}}
\tabletypesize{\tiny}
\tablehead{
\colhead{System ID} & \colhead{$z$} & \colhead{$\sigma_{\rm ap}$} & \colhead{m$_{8}$} & \colhead{r$_{\rm e8}$}    & \colhead{m$_{4}$}        & \colhead{r$_{\rm e4}$}}
\startdata
SDSS J003753.21$-$094220.1  & 0.1954 & 265 $\pm$ 10 & 15.75 $\pm$ 0.01 & 2.54 $\pm$ 0.02 & 18.85 $\pm$ 0.03 & 2.71 $\pm$ 0.07 \\
SDSS J021652.54$-$081345.3  & 0.3317 & 332 $\pm$ 23 & 16.29 $\pm$ 0.05 & 3.51 $\pm$ 0.22 & 19.74 $\pm$ 0.10 & 3.69 $\pm$ 0.37 \\
SDSS J073728.45$+$321618.5  & 0.3223 & 310 $\pm$ 15 & 16.47 $\pm$ 0.03 & 3.28 $\pm$ 0.13 & 19.92 $\pm$ 0.04 & 3.25 $\pm$ 0.12 \\
SDSS J091205.30$+$002901.1  & 0.1642 & 313 $\pm$ 12 & 15.03 $\pm$ 0.00 & 4.80 $\pm$ 0.02 & 17.92 $\pm$ 0.03 & 5.64 $\pm$ 0.17 \\
SDSS J095629.77$+$510006.6  & 0.2405 & 299 $\pm$ 16 & 16.18 $\pm$ 0.01 & 2.65 $\pm$ 0.03 & 19.58 $\pm$ 0.06 & 2.24 $\pm$ 0.14 \\
SDSS J095944.07$+$041017.0  & 0.1260 & 212 $\pm$ 12 & 16.34 $\pm$ 0.02 & 1.82 $\pm$ 0.05 & 19.11 $\pm$ 0.01 & 2.15 $\pm$ 0.02 \\
SDSS J125028.25$+$052349.0  & 0.2318 & 254 $\pm$ 14 & 16.22 $\pm$ 0.00 & 1.78 $\pm$ 0.01 & 19.53 $\pm$ 0.02 & 1.45 $\pm$ 0.02 \\
SDSS J133045.53$-$014841.6  & 0.0808 & 178 $\pm$  9 & 16.51 $\pm$ 0.00 & 1.23 $\pm$ 0.01 & 18.99 $\pm$ 0.01 & 1.41 $\pm$ 0.02 \\
SDSS J140228.21$+$632133.5  & 0.2046 & 275 $\pm$ 15 & 15.84 $\pm$ 0.00 & 3.12 $\pm$ 0.02 & 19.37 $\pm$ 0.08 & 2.33 $\pm$ 0.20 \\
SDSS J142015.85$+$601914.8  & 0.0629 & 194 $\pm$  5 & 14.63 $\pm$ 0.03 & 2.63 $\pm$ 0.10 & 17.13 $\pm$ 0.02 & 2.57 $\pm$ 0.05 \\
SDSS J162746.44$-$005357.5  & 0.2076 & 275 $\pm$ 12 & 16.27 $\pm$ 0.01 & 2.15 $\pm$ 0.02 & 19.71 $\pm$ 0.02 & 1.71 $\pm$ 0.03 \\
SDSS J163028.15$+$452036.2  & 0.2479 & 260 $\pm$ 16 & 16.34 $\pm$ 0.01 & 2.09 $\pm$ 0.02 & 19.79 $\pm$ 0.13 & 2.16 $\pm$ 0.31 \\
SDSS J230053.14$+$002237.9  & 0.2285 & 283 $\pm$ 18 & 16.57 $\pm$ 0.01 & 1.88 $\pm$ 0.01 & 20.05 $\pm$ 0.05 & 1.70 $\pm$ 0.09 \\
SDSS J230321.72$+$142217.9  & 0.1553 & 260 $\pm$ 15 & 15.23 $\pm$ 0.01 & 4.24 $\pm$ 0.04 & 18.21 $\pm$ 0.05 & 4.32 $\pm$ 0.19 \\
SDSS J232120.93$-$093910.2  & 0.0819 & 236 $\pm$  7 & 14.19 $\pm$ 0.00 & 4.49 $\pm$ 0.01 & 16.74 $\pm$ 0.01 & 5.58 $\pm$ 0.08 \\
\enddata
\tablecomments{For each SLACS lens (see coordinates and SDSS photometry in paper I) we list redshift, SDSS velocity 
dispersion, HST magnitude and effective radius through the F814W and
F435W filters (magnitudes are in the vega system, corrected for
Galactic Extinction, effective radii are in arcseconds).}
\end{deluxetable}

\clearpage
\begin{deluxetable}{lcccccccc}
\tablecaption{Summary of relevant corrected quantities \label{tab:rest}}
\tabletypesize{\tiny}
\tablehead{
\colhead{ID} & \colhead{$\sigma$} & \colhead{$\sigma_{\rm SIE}$} &\colhead{SB$_{\rm e,V}$} & \colhead{R$_{\rm e,V}$} & \colhead{\rm V}   & \colhead{SB$_{\rm e,B}$}        & \colhead{R$_{\rm e,B}$} & \colhead{\rm B}}
\startdata
SDSS J003753.21$-$094220.1 & 282$\pm$11 & 280 & 20.10$\pm$0.06 &  8.48$\pm$0.11 & -23.11 & 21.01$\pm$0.15 &  8.67$\pm$0.19 & -22.25  \\ 
SDSS J021652.54$-$081345.3 & 349$\pm$24 & 346 & 20.78$\pm$0.32 & 16.95$\pm$0.92 & -23.94 & 21.69$\pm$0.48 & 17.28$\pm$1.20 & -23.06  \\ 
SDSS J073728.45$+$321618.5 & 326$\pm$16 & 297 & 20.80$\pm$0.19 & 15.32$\pm$0.46 & -23.69 & 21.65$\pm$0.19 & 15.27$\pm$0.42 & -22.84  \\ 
SDSS J091205.30$+$002901.1 & 325$\pm$12 & 344 & 20.99$\pm$0.06 & 14.66$\pm$0.23 & -23.41 & 21.97$\pm$0.16 & 15.49$\pm$0.39 & -22.55  \\ 
SDSS J095629.77$+$510006.6 & 318$\pm$17 & 317 & 20.28$\pm$0.09 &  9.49$\pm$0.22 & -23.17 & 21.06$\pm$0.31 &  8.92$\pm$0.41 & -22.26  \\ 
SDSS J095944.07$+$041017.0 & 229$\pm$13 & 216 & 20.39$\pm$0.11 &  4.50$\pm$0.05 & -21.45 & 21.37$\pm$0.05 &  4.75$\pm$0.03 & -20.58  \\ 
SDSS J125028.25$+$052349.0 & 274$\pm$15 & 246 & 19.45$\pm$0.03 &  6.13$\pm$0.04 & -23.05 & 20.19$\pm$0.09 &  5.68$\pm$0.06 & -22.15  \\ 
SDSS J133045.53$-$014841.6 & 195$\pm$10 & 185 & 19.86$\pm$0.05 &  2.04$\pm$0.02 & -20.25 & 20.80$\pm$0.09 &  2.13$\pm$0.03 & -19.40  \\ 
SDSS J140228.21$+$632133.5 & 290$\pm$16 & 298 & 20.36$\pm$0.09 &  9.35$\pm$0.29 & -23.06 & 21.11$\pm$0.42 &  8.39$\pm$0.53 & -22.08  \\ 
SDSS J142015.85$+$601914.8 & 206$\pm$ 5 & 204 & 19.55$\pm$0.16 &  3.15$\pm$0.06 & -21.51 & 20.43$\pm$0.11 &  3.13$\pm$0.06 & -20.62  \\ 
SDSS J162746.44$-$005357.5 & 295$\pm$13 & 271 & 20.01$\pm$0.06 &  6.68$\pm$0.06 & -22.68 & 20.79$\pm$0.10 &  6.13$\pm$0.09 & -21.71  \\ 
SDSS J163028.15$+$452036.2 & 279$\pm$17 & 314 & 20.06$\pm$0.16 &  8.23$\pm$0.44 & -23.09 & 21.01$\pm$0.69 &  8.34$\pm$0.89 & -22.17  \\ 
SDSS J230053.14$+$002237.9 & 305$\pm$19 & 302 & 20.05$\pm$0.07 &  6.63$\pm$0.13 & -22.62 & 20.94$\pm$0.26 &  6.38$\pm$0.24 & -21.66  \\ 
SDSS J230321.72$+$142217.9 & 271$\pm$16 & 291 & 20.85$\pm$0.09 & 11.53$\pm$0.26 & -23.03 & 21.78$\pm$0.24 & 11.60$\pm$0.44 & -22.11  \\ 
SDSS J232120.93$-$093910.2 & 245$\pm$ 7 & 257 & 20.48$\pm$0.03 &  7.93$\pm$0.07 & -22.59 & 21.49$\pm$0.08 &  8.47$\pm$0.11 & -21.72  \\ 
\hline
\enddata
\tablecomments{For each SLACS lens as in Table~1 we list central
velocity dispersion (in \kms), velocity dispersion of the best fitting
singular isothermal ellipsoid, surface brightness, effective radius
and absolute magnitude in the B and V band rest frame (in mag
arcsec$^{-2}$ and kpc respectively). Uncertainties on the absolute
magnitude are equal to the uncertainties on the observed magnitudes,
plus approximately $\sim0.05$ mags to account for the K-color
correction.}
\end{deluxetable}

\clearpage

\end{document}